\documentclass[10pt,conference,a4paper]{IEEEtran}
\usepackage{times,amsmath,epsfig}
\usepackage{grffile}
\usepackage[bookmarks=false]{hyperref}
\usepackage{xcolor}

\providecommand*{\Heff}{\ensuremath{H_\mathrm{eff}}}
\providecommand*{\HGOE}{\ensuremath{H_\mathrm{GOE}}}

\providecommand*{\Hdirect}{\ensuremath{H_\mathrm{direct}}}
\providecommand*{\Hsys}{\ensuremath{H_\mathrm{system}}}
\providecommand*{\W}{\ensuremath{W}}
\providecommand*{\Smat}{\ensuremath{S}}
\providecommand*{\Trans}{\ensuremath{T}}
\providecommand*{\TA}{{\ensuremath{T_\mathrm{A}}}}

\providecommand*{\projdirect}[2]{\ensuremath{
    \left(\left\vert #1 \rangle\langle #2 \right\vert +
      \rule{0.0ex}{2.5ex}
      \left\vert #2 \rangle\langle #1 \right\vert\right)}}

\providecommand*{\ud}{\ensuremath{\mathrm{d}}}
\providecommand*{\ue}{\ensuremath{\mathrm{e}}}
\providecommand*{\ui}{\ensuremath{\mathrm{i}}}

\providecommand*{\mean}[1]{\ensuremath{\left\langle{#1}\right\rangle}}
\providecommand*{\Var}[1]{\ensuremath{\mathrm{Var}\left({#1}\right)}}
\providecommand*{\kappaA}{\ensuremath{\kappa_{\mathrm{A}}}}
\providecommand*{\lambdadp}{\ensuremath{\lambda}}
\providecommand*{\kappaAprime}{\ensuremath{\kappaA'}}
\providecommand*{\lambdaprime}{\ensuremath{\lambdadp'}}
\providecommand*{\optlambda}{\ensuremath{\lambdadp^{*}}}
\providecommand*{\optlambdanum}{\ensuremath{\optlambda_\mathrm{num}}}
\providecommand*{\abssq}[1]{\ensuremath{{
    \left\vert
      {#1}
    \right\vert^{2}}}}
\providecommand*{\mean}[1]{\ensuremath{
    {\langle}
      {#1}
    {\rangle}}}
\providecommand*{\var}[1]{\ensuremath{
    \mathrm{Var}\left({#1}\right)}}
\providecommand*{\ket}[1]{\ensuremath{
    {\left\vert {#1} \right\rangle}}}
\providecommand*{\bra}[1]{\ensuremath{
    {\left\langle {#1} \right\vert}}}
\providecommand*{\myfrac}[2]{\ensuremath{{#1}/{#2}}}

\title{Introducing Enhanced Transport to the Effective Hamiltonian
  Approach via Random Matrices with a Pair of Connecting States}

\author{%
  {
    Martin Richter{\small $~^{1}$},
    Fabrice Mortessagne,
    Olivier Legrand,
    Ulrich Kuhl{\small $~^{2}$}}%
\vspace{1.6mm}\\
\fontsize{10}{10}\selectfont\itshape
Institut de Physique de Nice, Universit\'{e} C\^{o}te d'Azur, CNRS, 06100 Nice, France\\
\fontsize{9}{9}\selectfont\ttfamily\upshape
$~^{1}$martin.richter@unice.fr
$~^{2}$ulrich.kuhl@unice.fr
}

\begin{document}
\maketitle
\begin{abstract}
  Direct transport processes play an important role in wireless
  communications where an ideal setup uses microwave fields to
  establish reliable communication channels between transmitter and
  receiver. But it is inherent to the problem that one cannot fully
  control the environment. While the influence of a complex scattering
  surrounding can be very well described using Random Matrix Theory it
  is not always obvious how to combine this universal approach with
  concrete communication channels. In this work we present an approach
  introducing an enhanced path between two antennas to the Hamilton
  operator to account for a prototypical problem. In order to be able
  to describe the stability of wireless chip-to-chip communication, we
  analyze the transport properties and predict the stability of the
  transmission under increasing importance of the environment.
\end{abstract}

\section{Introduction}
Using wireless communication is an ubiquitous technology in today's
life as it is the fundamental building block for mobile communications
as well as wireless computer networks. The progress of this technology
is reflected by increasing bandwidth and reliability as well as
reduced production costs, energy consumption and miniaturization.
These advances make it plausible to also think of using wireless
technology in places currently dominated by wired connections. One
of these fields is the very short range communication between chips
or even on integrated circuits.

While there have been various progresses in recent years on how to
improve the communication quality between WiFi
antennas~\cite{hen04,hou16,cou11}
this paper aims at another aspect of
microwave communications: How does the achieved quality of an antenna
setup, designed within an anechoic environment,
change if placed in a random, partly reverberating environment.
This is an important question as it is not
always possible to fully control the environment,
in particular inside computers or cell phones,
if one aims at plug-and-play solutions of antenna designs.

In this paper, we propose a physically motivated model and derive
analytical predictions in terms of the distribution of
transmission probabilities.
These results will be compared to numerical calculations.

One very common approach to model complexity in physical or even
sociological and financial systems relies on
random matrices in the framework
of Random Matrix Theory (RMT).
Starting from the claim that their
eigenvalues and eigenvectors show the same statistical properties as
systems with many degrees of freedom and intertwined
dynamics~\cite{stoe07b}, one finds astonishing agreements with a
wide range of experimental and numerical data given that the
described systems show a sufficient level of complexity~\cite{stoe07b}.

This allows to use these random matrices to derive analytical
expressions for measured, calculated or otherwise observed
quantities. However, one common aspect of the work with random
matrices when describing concrete problems is to incorporate all
non-universal aspects. This can either be done by incorporating them
into the theory or the other way around by removing all
non-universal features from the data and compare them with pure
RMT predictions.
One example of such non-universal features are direct reactions
in nuclear scattering. 

\section{Theoretical Model}
\label{sec:orgheadline10}

In order to describe the coupling of the established communication link
to the antennas we use an approach based upon the effective
Hamiltonian~\cite{stoe07b,mah69,sok89}
\begin{align}
\Heff = \Hsys - \frac{\ui}{2} \W\W^T.
\end{align}
This implies that the antennas themselves do not have any frequency
dependence and that there is a fixed number of transporting channels.
Furthermore, one assumes that all real shifts of the eigenvalues of
the closed system due to the opening can be absorbed into
\Hsys{} directly~\cite{kuh13}.
The communication between the antennas can be described by the
scattering matrix of the problem which is given by~\cite{mah69,sok89}
\begin{align}
  \Smat_{ij} = \delta_{ij} - \ui \left( \W^T\frac{1}{E -
  \Heff} \W \right)_{ij}.
\end{align}

Predictions based on this approach mostly assume that the underlying
complex system whose scattering properties are to be described has a
time-reversal symmetry and can therefore be modeled by an ensemble of
random matrices from the Gaussian Orthogonal Ensemble (GOE).
However, the predictions will only be valid if there are no further
non-universalities in the system.
Otherwise, the latter have to be taken into account appropriately.

One of the earliest approaches of these kinds have been concerned
with nuclear scattering experiments.
Here, the scattering can be divided into
fast-time scale processes by scattering at the total target
potential and slow-time scale processes exciting the target.
While RMT allows to describe the statistical properties of the
excitation, the direct reaction needs to be treated separately.
If a separation of time scales is possible, then this is usually
done by removing the fast-time scale process by means of scattering
phases of the scattering matrix~\cite{ver85a,nis85}.
One thereby removes the non-universal aspects from the system and
can compare the data to RMT predictions.
Such direct processes with shorter time scales might also be present
in microwave cavities and have to be accounted for as well,
for example by adapting the impedance
matrix~\cite{hem06}.

In terms of microwave transmission for WiFi communications, one cannot
straightforwardly use the same description for the direct
communication links between the antennas.
Indeed, the electromagnetic field in the vicinity of printed circuit
boards (PCBs), in the presence of external noise due to other computer
components, or to an unknown environment, does not necessarily lead to a
separation of time-scales in the signals.

This paper addresses this issue by creating a model Hamiltonian which
allows for a description of an inter-antenna transmission whose time
scale cannot be well separated from that of the background.

\section{Model Hamiltonian}

In order to model a direct process between two antennas, we choose the
following coupling matrix \W{}~\cite{leh95a}
\begin{align}
  \W^T & = \sqrt{\frac{2\kappaA\Delta}{\pi}}\sqrt{N}\begin{pmatrix}
    1 & 0 & 0 & \cdots\\
    0 & 1 & 0 & \cdots
  \end{pmatrix},
\end{align}
where \(N\) is the size of the Hamiltonian, \(\Delta\) its mean level
spacing, and \kappaA{} is the antenna coupling strength related to the
so-called antenna transmission \TA{} by the relation~\cite{ver85a}
\begin{align}
  1 - \abssq{\mean{\Smat_{ii}}} =:
  \TA(\kappaA) =
  \frac{4 \kappaA}{\abssq{1 +
  \kappaA}}
\end{align}
This choice allows to identify the channels of the scattering system
(rows in \W) directly with the first two basis vectors
in which the Hamiltonian
\(\Hsys{} = \sum_{nm}\ket{n}H_{nm}\bra{m}\)
is represented.
In the following, the mean level spacing is chosen to be unity \(\Delta
= 1\).
We model the direct process by a dyadic operator:
\begin{align}
  \label{eq:direct-process-def}
  \Hdirect &= \frac{\sqrt{N}}{\pi} \lambdadp\projdirect{1}{2}.
\end{align}
In the absence of any chaotic back reflections from the environment, the
system Hamiltonian reads \(\Hsys = \Hdirect + \frac{\ui}{2}\W\W^T\) 
and can be reduced to its upper \(2\times 2\) block.
We can straightforwardly derive the scattering matrix at \(E=0\)
\begin{align}
  \Smat = 1 + \frac{2\ui \kappaA}{\kappaA^2 +
   \lambdadp^2 / N}\begin{pmatrix} \ui \kappaA & \lambdadp / \sqrt{N} \\
   \lambdadp / \sqrt{N} & \ui \kappaA\end{pmatrix}
\end{align}
which leads to a transmission of
\begin{align}
  \Trans = |\Smat_{12}(E=0)|^2 =
  T_A\left(
  \frac{N\kappaA^2}{\lambdadp^2}\right)
\end{align}
with the already introduced function \TA. This leads to a distribution
of transmissions
\begin{align}
  \label{eq:transmission-delta-distribution}
  P(\Trans) = \delta\left(\Trans - T_A\left(
  \frac{N\kappaA^2}{\lambdadp^2}\right)\right)
\end{align}
which leads to a maximal transmission of \(\Trans = 1\) at an optimal
parameter \optlambda{}
\begin{align}
  \label{eq:optimal-lambda}
  \optlambda_0(\kappaA) = \sqrt{N}\kappaA.
\end{align}

If we use the above \Hdirect{} from Eq.~\eqref{eq:direct-process-def} in
order to include the enhanced transmission into the complex environment,
we use the dyadic operator augmented by a GOE Hamiltonian
\begin{align}
  \label{eq:model-GOE-and-direct}
  \Hsys = \Hdirect + \HGOE.
\end{align}
In the following, we will elucidate how this changes the unperturbed
result~\eqref{eq:transmission-delta-distribution}.
The elements \(h_{ij}\) of \HGOE{} are zero-mean Gaussian
random variables with
\begin{align}
  \label{eq:GOE-variance}
  \langle h_{ij} h_{kl} \rangle =
  \left\{\begin{array}{rl}
          \frac{N}{\pi^2}\delta_{ik}\delta_{jl}
          & i \neq j \\[1ex]
          2
          \frac{N}{\pi^2}\delta_{ik}\delta_{jl}
          & i = j \end{array}\right.
\end{align}
which ensures that the mean density of states at zero energy is one and
therefore \(\Delta = 1\) as mentioned above.
This choice explains the prefactor in
Eq.~\eqref{eq:direct-process-def} which corresponds to the standard
deviation of the elements of \HGOE{}.

The general dependence of the distribution of the transmissions
\(P(\Trans)\) can be qualitatively read off from calculating the mean
transmission, \(\mean{\Trans}\) vs. both relevant parameters,
\kappaA{} and \lambdadp.
\begin{figure}[t]
  \centering
  \includegraphics{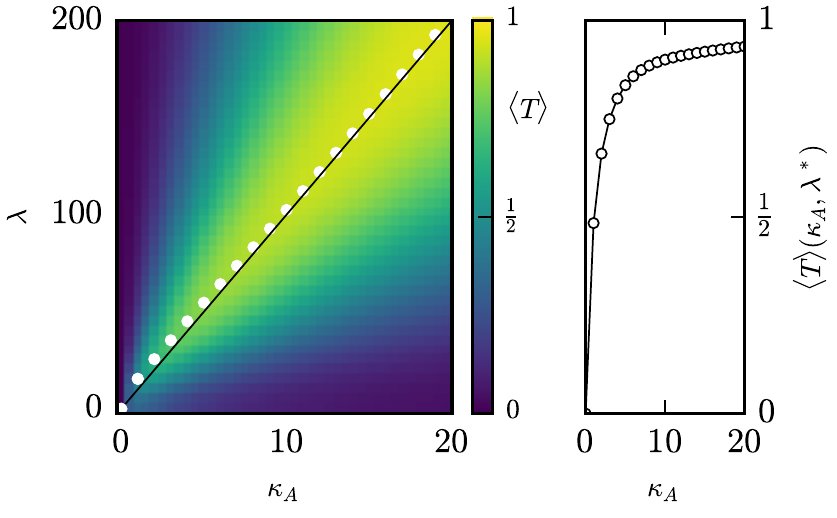}
  \vspace*{-2ex}
  \caption[Average transmission]{%
    Left panel: %
    Parameter dependence of the average
    transmission \(\mean{\Trans}(\kappaA, \lambdadp)\).
    The white dots are the optimal values of
    \(\optlambdanum\) obtained by a numerical search for
    fixed \kappaA.
    The straight black line is given by Eq.~\eqref{eq:optimal-lambda}
    for comparison.
    Right panel: %
    Average transmission \(\Trans(\kappaA, \optlambdanum)\)
    for the numerically obtained
    optimal parameters.
    The calculation at each point \(\kappaA, \lambdadp\) is based on
    \(80\) realizations of \(100 \times 100\) GOE Hamiltonians.
  }
  \label{fig:trans-dependency-kappa-lambda}
\end{figure}

\begin{figure}[t]
  \centering
  \includegraphics[width=1.0\columnwidth]{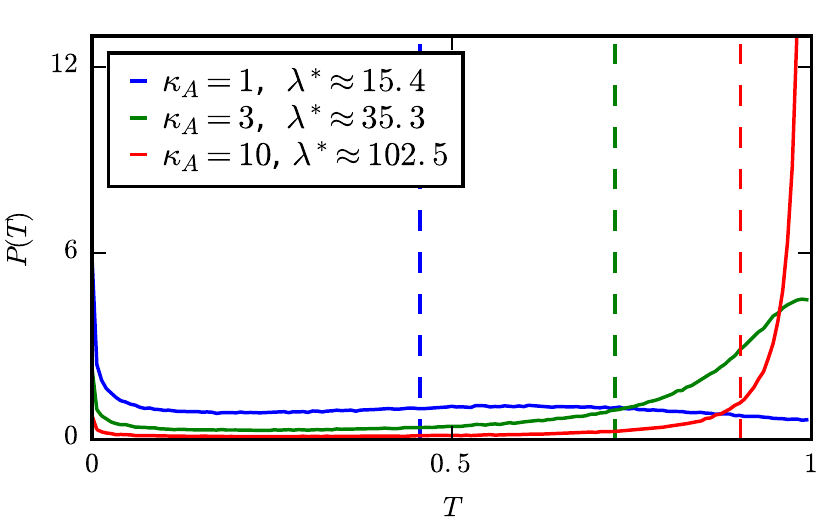}
  \vspace*{-1ex}
  \caption[Transmission Distributions]{
    Transmission distribution for three pairs of \(\kappaA, \lambdadp\)
    values from the numerically obtained optimal curve for maximal
    average transmission \mean{\Trans}.
    The curves belong to increasing value of \kappaA{} for blue,
    green, and red.
    The dashed lines indicate \(\mean{T}\) for every curve.
    The plot emphasizes, that also the variance of the distribution
    gets smaller for larger values of the parameters.
    For these curves \(8\cdot 10^{5}\) realizations have been used.
  }
  \label{fig:trans-distributions-for-optimal-params}
\end{figure}

A corresponding plot is shown in
Fig.~\ref{fig:trans-dependency-kappa-lambda}.
From the figure, one can see that the numerically obtained optimal value
\(\optlambdanum(\kappaA)\) does tend towards the straight line given by
the case without environment in Eq.~\eqref{eq:optimal-lambda}.
The same result is obtained when not optimizing numerically
for maximal \(\mean{T}\) but for minimal \(\Var{T}\) (not shown).
At small values of \(\kappaA\) the dependency deviates from the linear
form.
However, the interesting parameter regime is at large
values as this region corresponds to large \mean{\Trans}, as can
be seen from the right panel in
Fig.~\ref{fig:trans-dependency-kappa-lambda}.
In order to check whether the region
\begin{align}
  \label{eq:interesting region}
  \kappaA \gg 1 \qquad \mathrm{and} \qquad
  \lambdadp \approx \optlambda \gg 1
\end{align}
is the interesting parameter range, we present three distributions of
transmissions along the line in
Fig.~\ref{fig:trans-distributions-for-optimal-params}.

In order to get a smooth dependency of \(P(\Trans)\), we used
\(8\cdot 10^{5}\) realizations for each of the curves.
The plot clearly shows that the distribution \(P(\Trans)\) gets closer
to the expression~\eqref{eq:transmission-delta-distribution} for
large values of \kappaA{} and \lambdadp.

\section{Perturbation Theory}

As the most promising parameter region for high transmission can be
found for large parameter values~\eqref{eq:interesting region}, we can
approach the complex background perturbatively.
We can do so by introducing a small parameter \(\varepsilon\) in
\begin{align}
\lambdadp = \frac{\lambdaprime}{\varepsilon}
  \hspace{3ex}
  \kappaA = \frac{\kappaAprime}{\varepsilon}
  \hspace{3ex}
  \Heff' = \varepsilon \Heff
\end{align}
which scales the parameters up to large values.
With this parametrization, the GOE
contribution of the Hamiltonian becomes a small
perturbation.
With
\begin{align}
  \label{eq:perturbative-Heff}
  \Heff' = \begin{pmatrix}
    -\ui\frac{\kappaAprime}{\pi}N &
    \lambdaprime\frac{\sqrt{N}}{\pi} & 0 & \\
    \lambdaprime\frac{\sqrt{N}}{\pi} &
    -\ui\frac{\kappaAprime}{\pi}N
    & 0 & \dots
    \\[1ex]
    0 & 0 & 0 & \\
    & \vdots & & \ddots \\
  \end{pmatrix} + \varepsilon
  \HGOE
\end{align}
we can first diagonalize the non-GOE part.
The corresponding eigenvalues and eigenvectors are
\begin{align}
  \left(\Hdirect + \frac{\ui}{2}\W\W^T\right)
  \ket{r^{(0)}_n}  = E^{(0)}_n \ket{r^{(0)}_n}.
\end{align}

Although this Hamiltonian is
non-hermitian, its symmetric form has right eigenvectors which are equal
to the left eigenvectors, \(\ket{l^{(0)}} = \ket{r^{(0)}}\).
Under the presence of \(\varepsilon \HGOE\) we can write the eigenenergies
and eigenvectors
in first order perturbation theory for non-hermitian
operators as~\cite{fyo12}
\begin{align}
  \label{eq:first-order-evals-evecs}
  \tilde{E}^{(1)}_n &= E^{(0)}_n + \varepsilon E^{(1)}_n,
  &
  \ket{\tilde{r}^{(1)}_n} &= \ket{r^{(0)}_n} + \epsilon
  \ket{r^{(1)}_n}
\end{align}
and arrive at
\begin{align}
  \Smat_{12} =
  \ui \frac{\kappaAprime\Delta}{2\pi}N
  & \left(\bra{r^{(0)}_2}-\bra{r^{(0)}_1}\right)
  \nonumber
  \\
  &
    \sum\limits_{n=1}^{2}\ket{\tilde{r}^{(1)}_n}
    \frac{1}{\varepsilon E - \tilde{E}^{(1)}_n}
    \bra{\tilde{r}^{(1)}_n}
  \\
  \nonumber
  &
    \left(\ket{r^{(0)}_1} + \ket{r^{(0)}_2}\right).
\end{align}
Because all eigenvalues \(E_n^{(0)}\) of the unperturbed
Hamiltonian~\eqref{eq:perturbative-Heff} with \(n > 2\)
are degenerate, the corresponding eigenvectors up to
first order are just mixed among each other.
Therefore, they are orthogonal to the vectors
\(\ket{r^{(0)}_1} \pm \ket{r^{(0)}_2}\) and do not contribute
to \(\Smat_{12}\).
Hence, only the first two elements of the sum remain thereby
effectively reducing the \(N(N + 1)/2\) independent parameters
from the \(N \times N\) \HGOE{} to only \(h_{11}, h_{12}\), and \(h_{22}\).
After some algebra we can write
\begin{align}
  \Smat_{12}    = -2\mathrm{i}\left(1 - z^2\right)
  \frac{y} {(x + \mathrm{i})^2 - y^2}
\end{align}
and therefore
\begin{align}
  \label{eq:S12-abssqr}
  \abssq{\Smat_{12}}
  &= \left(1 - z^2\right)^2\frac{4 y^2}{
    \left(x^2 - y^2 - 1 \right)^2 + 4x^2}
  \\
  &=: \left(1 - z^2\right)^2 c(x, y)
\end{align}
where we used the abbreviations
\begin{align}
  \label{eq:definition-xyz}
  x
  &= -\frac{h_a}{\kappaA\sqrt{N}}
  &
    y &= \frac{\lambdadp}{\kappaA\sqrt{N}} +
        \frac{h_c}{\kappaA\sqrt{N}}
  &
  z &= \frac{h_b}{2\lambdadp}.
\end{align}
In these expressions the variables \(h_{a,b} := \pi (h_{11} \pm h_{22}) /
2 \sqrt{N}\) and \(h_c := \pi h_{12} / \sqrt{N}\) are
normal-distributed \(\mathcal{N}(0, 1)\) random variables.
It follows that the variances read
\begin{align}
  \label{eq:variances-xyz}
  \var{x} = \var{y} &= \frac{1}{\kappaA^2N}
  &
    \var{z} &= \frac{1}{4\lambdadp^2}.
\end{align}

With this conventions our main result follows as
\begin{align}
  P(T)
  =& \mean{\delta\left(T - \abssq{S_{12}}\right)} \\
  =&\int\!\!\!\!\int\!
    \ud {h_a}
    \frac{\ue^{-h_a^2/2}}{\sqrt{2\pi}}
    \ud {h_c}
    \frac{\ue^{-h_c^2/2}}{\sqrt{2\pi}}
    \!\frac{\lambdadp / c}{\sqrt{2\pi}}
    \frac{1}{\sqrt{\myfrac{\Trans}{c}}}
    \label{eq:perturbative-result}
  \\
  &
    \nonumber
    \left(\!
    \frac{\ue^{-2\lambdadp^2\left(1 + \sqrt{\myfrac{\Trans}{c}}\right)}}{
    \sqrt{1 + \sqrt{\myfrac{\Trans}{c}}}}
    \!+\!\Theta\left(c - \Trans\right)
    \frac{\ue^{-2\lambdadp^2\left(1 - \sqrt{\myfrac{\Trans}{c}}\right)}}{
    \sqrt{1 - \sqrt{\myfrac{\Trans}{c}}}}
    \!\right).
\end{align}

The above result can be further simplified when considering the fact
that the variances in Eq.~\eqref{eq:variances-xyz} go to zero in the
interesting limit~\eqref{eq:interesting region}.
This allows to evaluate the Gaussians only at their peak.
This leads us to
\begin{align}
  c\left(0, \frac{\lambdadp}{\kappaA\sqrt{N}}\right) =
  \TA\left(\frac{\lambdadp^2}{N \kappaA^2}\right)
\end{align}
and therefore to
\begin{align}
  \label{eq:steepest-descent}
  P(T)\,\approx\,
  &
    \frac{\lambdadp / \TA}{\sqrt{2\pi}}
    \frac{1}{\sqrt{\myfrac{\Trans}{\TA}}}
    \left(\rule[0ex]{0ex}{5ex}\right.
    \frac{\ue^{-2\lambdadp^2\left(1 + \sqrt{\myfrac{\Trans}{\TA}}\right)}}{
    \sqrt{1 + \sqrt{\myfrac{\Trans}{\TA}}}}
    \nonumber
  \\
  &
    + \Theta\left(\TA - \Trans\right)
    \frac{\ue^{-2\lambdadp^2\left(1 - \sqrt{\myfrac{\Trans}{\TA}}\right)}}{
    \sqrt{1 - \sqrt{\myfrac{\Trans}{\TA}}}}
    \left.\rule[0ex]{0ex}{5ex}\right).
\end{align}
Note that this treatment is only approximately correct.
A proper use of
the method of steepest-descent yields a more complicated point for the
saddle as it is shifted by
the exponential contributions inside the brackets.
However, the actual shift turns out to be negligible.

\section{Numerical Results}

\begin{figure}[t]
  \centering
  \includegraphics[width=0.95\columnwidth]{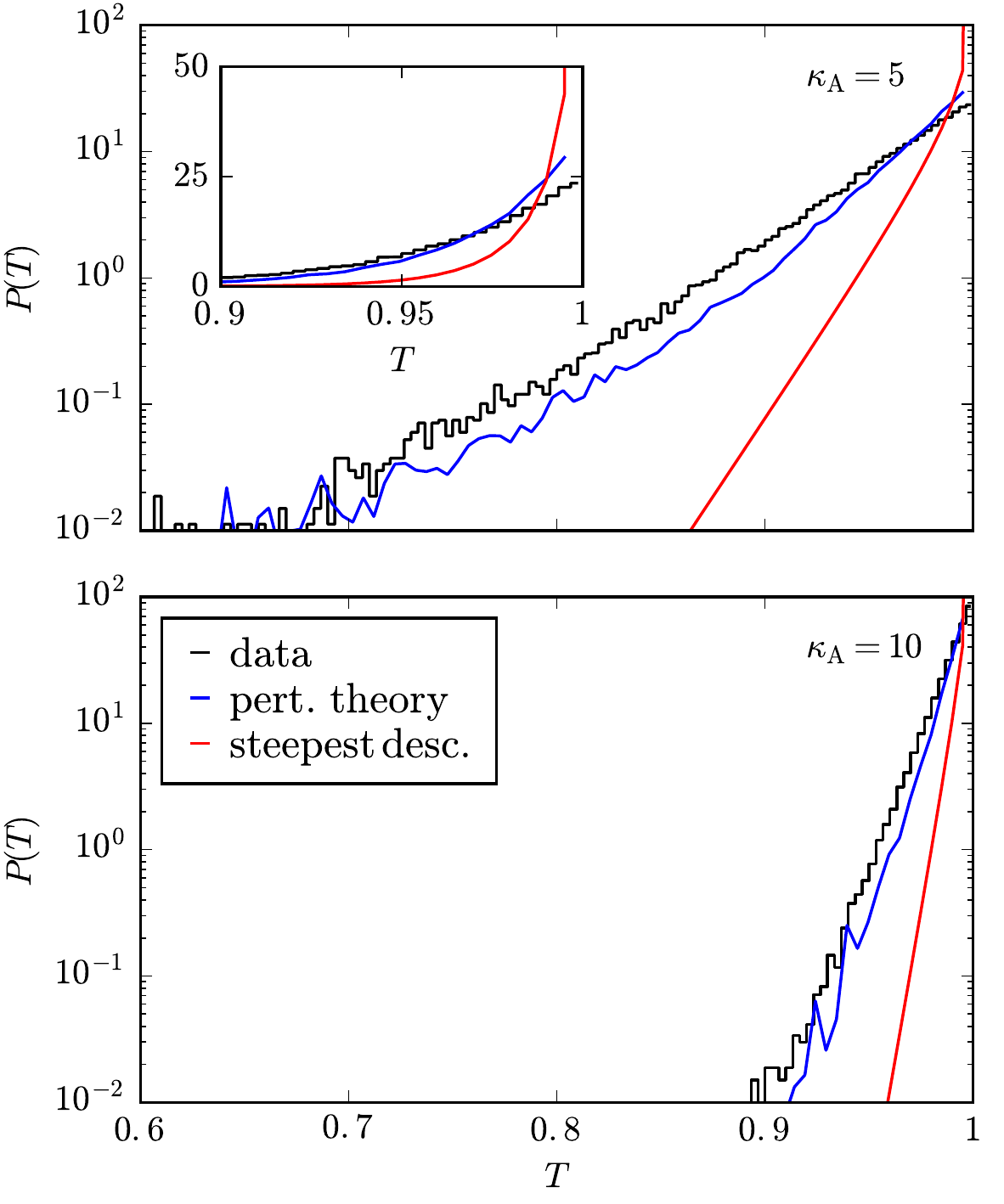}
  \vspace*{-1ex}
  \caption{Perturbative prediction of \(P(\Trans)\)
    for \(\kappaA = 5\) (upper panel) and \(\kappaA = 10\) (lower panel).
    Shown in black are RMT calculations from \(2\times 2\) Hamiltonians
    for the optimal \lambdadp{} from Eq.~\eqref{eq:optimal-lambda}.
    The calculation was carried out for \(8\cdot 10^3\) realizations.
    Shown in blue is the perturbative
    result~\eqref{eq:perturbative-result} while red is the
    steepest-descent approximation~\eqref{eq:steepest-descent}.
    The upper panel contains an inset in linear scale for comparison.
  }
  \label{fig:trans-dependency-kappa-lambda-perturb}
\end{figure}

In order to check the range of validity of the resulting
formulas~\eqref{eq:perturbative-result} and~\eqref{eq:steepest-descent},
we can compare them to RMT calculations as done previously in
Fig.~\ref{fig:trans-distributions-for-optimal-params}.
The comparison is shown in Fig.~\ref{fig:trans-dependency-kappa-lambda-perturb}
for two values of the parameter \(\kappaA = 5, 10\).
While the perturbative results reasonably match the RMT data, the
steepest-descent solution proves to be too crude to fully estimate the
width of \(P(\Trans)\).
Note that due to the optimal choice of
\lambdadp{}, Eq.~\eqref{eq:optimal-lambda}, the distribution peaks at
\(\Trans = 1\).
In comparison, the solution without any complex background,
Eq.~\eqref{eq:transmission-delta-distribution}, is a \(\delta\)
peak exactly at this value, \(P(\Trans) = \delta(\Trans - 1)\).

\section{Conclusions}

This paper presents a physically motivated effective Hamiltonian
approach to a common problem of microwave communications.
The propagation of electromagnetic fields over PCB boards in the
vicinity of obstacles does not allow to use a separation of time scales.
This excludes common approaches to account for direct communication
paths and makes new models, as the one presented, necessary.
By including the direct communication directly into the Hamiltonian, the
model allows to determine relevant parameter ranges and a perturbative
description of the distribution of transmissions \(P(\Trans)\).
The resulting approximately exponential decay provides an expression
which can be easily compared with experimental data for example from
chaotic reverberation chambers (CRCs).
In subsequent works we will extract the corresponding parameters from
measured transmission and reflection spectra.
The above formulas give then rise to distinguish between the stability
of several antenna setups (patch, monopole, horn) and give design
guidelines for stable communications on small scales.

\section*{Acknowledgment}
\label{sec:orgheadline24}

This work was supported in part by the European Union Horizon 2020
research and innovation program under grant no. 664828
(NEMF21~\cite{nemf21}).

\end{document}